\documentstyle[twocolumn,prl,aps, epsfig]{revtex}
\begin{document} 
\title{
Meta-Percolation and Metal-Insulator Transition in Two Dimensions
}
\author{Junren Shi$^{1}$, Song He$^{2}$, and X. C. Xie$^{1}$}
\address{
1. Department of Physics, Oklahoma State University,
Stillwater, OK 74078
}
\address{
2. Hexaa Laboratory, Warren, NJ 07059
}

\address{\rm (submitted to Phys. Rev. Lett. on March 8, 1999)}
\address{\mbox{ }}
\address{\parbox{14cm}{\rm \mbox{ }\mbox{ }
According to the scaling theory of localization, 
all quantum electronic states are
localized in two-dimensional (2D)
systems. One consequence of the theory is that there is
no quantum percolation transition in 2D.
However, in a real system at a finite temperature,
electron phase coherent length is finite and the system is
between quantum and classical. We find, in such a 2D system,
a metal-insulator transition (MIT) caused by a novel type
of percolation, {\it meta-percolation}.
The relevance to recently observed 2D MIT is also discussed.
}}
\address{\mbox{ }}
\address{\parbox{14cm}{\rm PACS numbers: 71.30.+h, 73.40.Hm}}
\maketitle


\vspace{-0.5cm}

There have been long lasting interests to understand localization problem in
two-dimensional (2D) electron systems.
According to the scaling theory of localization\cite{Abrahams,Lee}, there is no
metal-insulator transition in non-interacting two-dimensional (2D)
electron systems. One consequence of the theory is that there is
no 2D quantum percolation transition. On the other hand, it is well
known that there exists 2D
classical percolation transition.
Thus, it is interesting to understand that in a realistic electron system,
where electron phase coherence length is finite, whether the system is
quantum like, or classical like, or even more, unlike either, belongs to
a new class.

In this paper, we study a 2D quantum percolation model\cite{Shapir} with
finite phase coherence length. The dephasing mechanism is introduced by
attaching current-conserving voltage leads to the system, a method widely
used for one-dimensional models in mesoscopic community\cite{Buttiker,Datta}.  
The conductance in such a system is derived and is found to contain both
quantum and  classical contributions. The conductance $g$ for a finite size
system is calculated. Through the finite size scaling analysis,
we show clear evidence of metal-insulator transition (MIT). The MIT is
driven by a novel type percolation transition
which is semi-quantum and semi-classical in nature. We call it meta-percolation.
Finally, we discuss the
possible relevance to the newly observed 2D MIT at zero magnetic field.

\begin{figure}
\vspace{2mm}
\epsfig{file=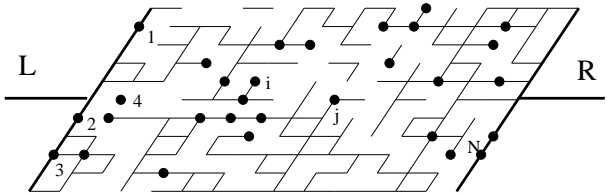, width=0.45\textwidth, clip}
\vspace{2mm}
\caption{
Lattice system of our study. $\bullet$ denotes a lattice site
where a phase-breaking voltage lead is attached.
\label{fig1}}
\end{figure}
                                                                                  
Our quantum percolation model\cite{Shapir} is based on a tight binding
Hamiltonian of a 2D square lattice,
\begin{eqnarray}  
{\cal H} = \sum_i \epsilon _{i}c_{i}^{+}c_{i}+ \sum_{<i,j>}t_{ij} 
(c_{i}^{+}c_{j}+c_{j}^{+}c_{i}),
\end{eqnarray}
where $<ij>$ denotes a pair of nearest-neighbor sites and
$\epsilon_{i}$ is the  on-site energy. We choose $\epsilon_{i}$ to
be random, ranging from $-W/2$ to $W/2$ to model an
on-site disorder. The nearest-neighbor hopping
matrix elements $t_{ij}$ is a random variable which assumes the
values $1$ (connected bonds in Fig.1)
or $0$ (empty bonds in Fig.1) with respective probabilities $P$ and $1-P$.
In order to introduce phase-breaking mechanism, we attach current-conserving
voltage leads\cite{Buttiker,Datta} 
at random lattice sites ($\bullet$ sites in Fig.1). 
The probability
where a lattice site is attached with a voltage lead is denoted
$P_{v}$ and the hopping element between a voltage lead and a lattice
site is denoted $t_{v}$. The sole role of the voltage leads is to
randomize the phases of incoming and outgoing wave-functions at the lead
sites while maintain current conservation.
Using Keldysh Green's function formalism\cite{Datta,Haug}, 
it is straightforward to derive\cite{Shi} 
the multi-lead current-voltage relation:
\begin{eqnarray} 
I_{i}=\sum _{i=1}^{N}\sigma _{ij}V_{j}\, \, \, \, i,\, 
j=L,\, R,\, 1,\, 2,\, \cdots \, N, 
\end{eqnarray}
where
\begin{eqnarray}
\sigma _{ij}=\frac{8e^{2}}{\hbar }Tr\left[ (T_{j\alpha }G^{r}_{\alpha \alpha }T_{\alpha i})\rho _{i}(T_{i\alpha }G^{a}_{\alpha \alpha }T_{\alpha j})\rho _{j}\right] \, \, \, \, \, \, \, i\neq j 
\end{eqnarray}
and
\begin{eqnarray}
\sigma _{ii}=-\sum _{j\neq i}\sigma _{ij}
\end{eqnarray}
In the above equations, 
$\alpha$ denotes the vector
space of the lattice system, $T_{\alpha i}$ is the transmission matrix, 
$G^{r(a)}_{\alpha \alpha }$ is the
retarded (advanced) Green's function of the lattice system, and
$\rho_{i}=Im g_{i}^{a}$ are the
density of states matrices for the voltage leads and measurement
leads $L$ and $R$, with $g_{i}^{a}$ being the advanced Green's function
at $i$th lead. In order to satisfy current conservation, we
require that the total current through each voltage lead to be
zero, namely, $I_{i}=0$ when $i\neq L, R$. This restriction fixes the
values for $V_{i}$. Under these conditions,
we obtain the conductance between $L$ and $R$
\begin{eqnarray}
g=\sigma _{LR}^{d}+\sigma _{LR}^{i}.
\end{eqnarray}
Here $\sigma _{LR}^{d}$ is the direct conductance which can be
obtained from Eq.(3) with $i=L$ and $j=R$, and
$\sigma _{LR}^{i}$ is the indirect
conductance,
\begin{eqnarray}
\sigma _{LR}^{i}  =   
-\textrm{Tr} & & \left\{ \left[ \begin{array}{ccc}
\sigma _{L1} & \cdots  & \sigma _{LN}\\
\sigma _{R1} & \cdots  & \sigma _{RN}
\end{array}\right] \left[ \begin{array}{ccc}
\sigma _{11} & \cdots  & \sigma _{1N}\\
\vdots  & \ddots  & \vdots \\
\sigma _{N1} & \cdots  & \sigma _{NN}
\end{array}\right] ^{-1}\right. \nonumber \\
& &
\left.  \left[ \begin{array}{cc}
\sigma _{1L} & \sigma _{1R}\\
\vdots  & \vdots \\
\sigma _{NL} & \sigma _{NR}
\end{array}\right] \right\}.
\end{eqnarray}
The $\sigma _{LR}^{i}$ is the conductance of the conductance network
obtained from the quantum lattice model.
We should mention that although the expression for the
direct conductance $\sigma _{LR}^{d}$
is identical for the pure quantum case without phase-breaking
voltage leads, its value is different with voltage leads attached since
the Green's functions in the formula are affected by the presence
of voltage leads.


We have calculated the total conductance $g$ numerically. In the following
we present the results and discuss their implications.
There are several parameters in our model as mentioned before:
$P$ (probability for $t_{ij}=1$), $P_{v}$ (probability for
a voltage lead), $t_{v}$ (hopping between a voltage lead and the lattice
system), $W$ (on-site disorder), and $E$ (electron
energy). The results presented below are for
$t_{v}=0.1$, $W=2$ and $E=0$. We have done systematical studies by varying 
these three parameters, including zero on-site disorder with $W=0$, and we
found that the qualitative results remained unchanged.
In Fig.2(a) we show $g$ as a function of $P$ for a pure
quantum case with $P_{v}=0$. The calculation is done on an $L \times L$
square lattice. The different curves are for different $L$'s with
$L=10$ (solid line), $L=20$ (dashed line) and $L=30$ (long-dashed line).
The important message from this plot is that
for a given $P$, $g$ decreases with increasing system size $L$,
indicating an insulating behavior. This result is consistent with the
conclusion of scaling theory of localization, namely, in a
disordered quantum system, all states are localized. 

  In Fig.2(b)
we plot the same curves as in Fig.2(a), except with $P_{v}=0.2$,
{\it {i.e.}}, $20\%$ of lattice sites are attached with phase-breaking
voltage leads. The striking difference between the two plots is
that in Fig.2(b), all curves cross at a single point $P_{c}\simeq 0.62$.
For $P>P_{c}$, $g$ increases with $L$, while it decreases
with $L$ for $P<P_{c}$. Thus, the system is metallic above $P_c$
and insulating below $P_c$ with a MIT at $P=P_{c}$. At $P_c$, $g$
is independent of $L$, signifying a divergent length $\xi$ there.
Therefore, the MIT is a second-order phase transition.

\begin{figure}
\vspace{2mm}
\epsfig{file=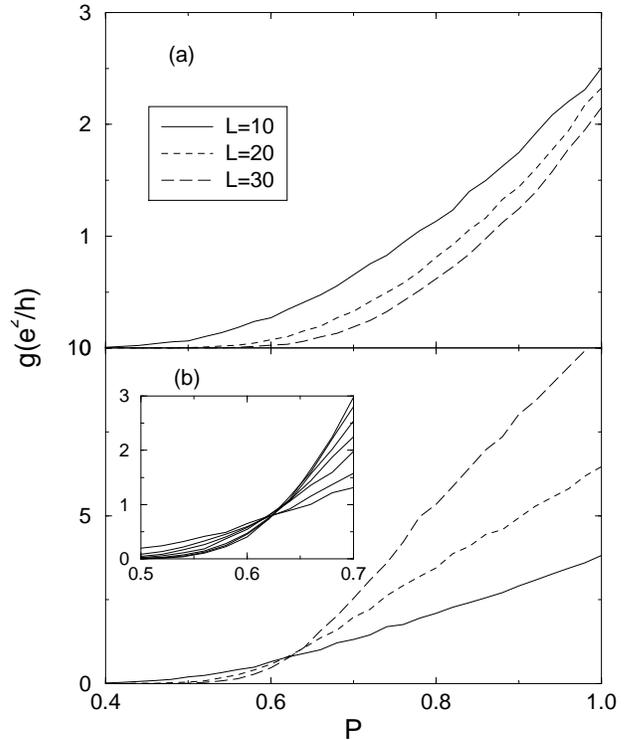, width=0.45\textwidth, clip}
\caption{
(a) Conductance $g$ as a function of the probability $P$ for a pure
quantum system with size $L=10$, $20$ and $30$.
(b) Same as (a), except with $P_{v}=0.2$.
Inset: Plot $g$ vs. $P$ at the vicinity of $P_{c}$ for
$L=10$, $15$, $20$, $25$, $30$, $35$ and $40$.
\label{fig2}}
\end{figure}
 
What are the nature of the transition and
the physical meaning of $\xi$?
The phase coherent length
$L_\phi$ defines a cut-off length scale beyond which the classical laws for
calculating a conductance are valid. One can divide a system into 
blocks with area $L_{\phi} \times L_{\phi}$. Within each block, the conductance
has to be determined quantum mechanically, {\it i.e.}, localization physics
plays an important role. The total conductance 
of the system is obtained by applying the classical laws to
the block conductances.  This defines a random
conductance network problem. The localization length $\lambda_{loc}$
of the 2D quantum system, which is always finite,
determines the average conductance and conductance fluctuations of each
block. For the quantum percolation model studied in this work, 
$\lambda_{loc}$ depends on the probability $P$ for a fixed on-site disorder $W$.
At a small $P$ such that $\lambda_{loc} \ll L_{\phi}$, the block
conductance $g_{block}\sim e^{-L_{\phi}/\lambda_{loc}}$ and the system is an
insulator. With increasing $P$ such that $\lambda_{loc} \simeq L_{\phi}$,
the average block conductance $\overline{g}_{block}$ is of the order
$e^{2}/h$. The detail distribution of $g_{block}$ is determined by
the corresponding quantum Hamiltonian. For the quantum percolation model,
$g_{block}=0$ if inside the block there is no at least one connected
quantum path within which every nearest-neighbor hopping $t_{ij}=1$. Hence,
the distribution function of $g_{block}$ always has a finite
$\delta$-function like peak at $g_{block}=0$, while it is a continuous
function for $g_{block}>0$. This defines a classical percolation
problem in which $g_{block}$ may be zero or falls into a continuous
distribution function.
In this new conductance network, there exists a
characteristic length $\xi$, which is the average size of the connected
cluster. $\xi$ diverges at $P=P_{c}$, responsible for the fix point
in Fig.2(b). 

\begin{figure}
\vspace{2mm}
\epsfig{file=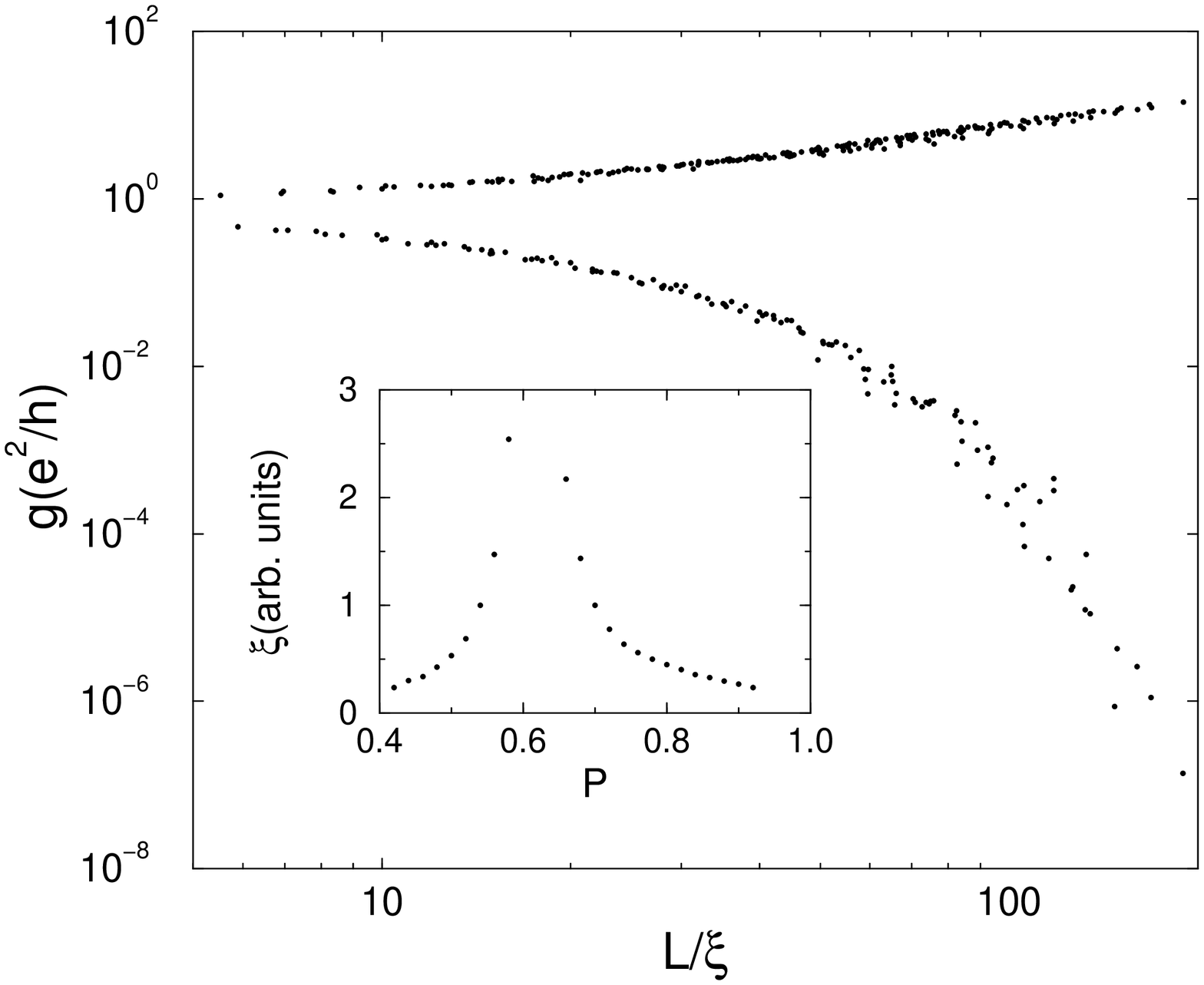, width=0.35\textwidth, angle=270}
\epsfig{file=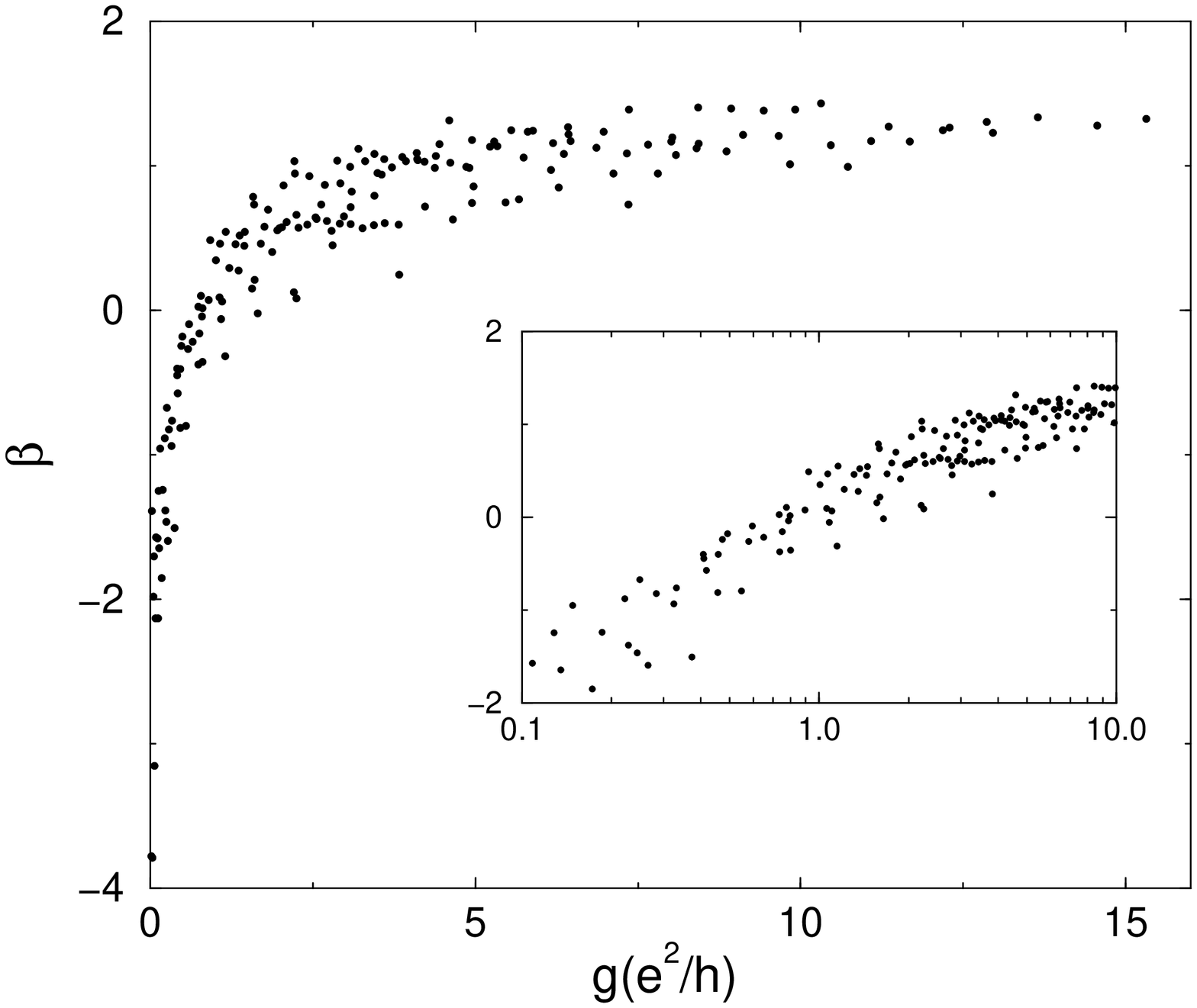, width=0.35\textwidth, angle=270}
\caption{
(a) Scaling plot of $g$ vs. $L/\xi$ for
$L=10$, $12$, $15$, $18$, $20$, $20$, $25$, $28$, $30$,
$32$, $35$, $38$, $40$ and $45$.
Inset: $\xi$ vs. the probability $P$.
(b) Plot $\beta$ function vs. $g$. Inset: $\beta$ vs. $\log g$
near $\beta =0$.
\label{fig3}}
\end{figure}                                           

  In Fig.3 we show the scaling properties of our data.
We use the standard finite-size scaling analysis\cite{Barber}.
The conductance curves for system sizes $L$ and different $P$'s can be
scaled to the metallic and insulating branches. Those curves with $P>P_c$
($P_c=0.62$) on the metallic side
are scaled along the $L$-axis so that they coincide
with a particular curve chosen here for $P=0.70$.
$\xi(P)$ is determined from curve shifting.
The similar procedure is also applied to insulating side,
in which case we scaled the curves to coincide with the curve for $P=0.54$.
We find that
all the conductance $g$ for different size $L$
(ranging from $10$ to $45$) can be
collapsed in a two-branch scaling function,
$g(L)=f(L/\xi(P))$, where $\xi(P)$ diverges at $P_{c}$.
Fig.3(a) shows the scaling function and $\xi(P)$ is shown in the inset.
We find that $\xi =|P-P_{c}|^{-\nu}$ with $\nu =1.4 \pm 0.2$.
On the insulating side, to a very good approximation,
$g \propto e^{-L/\xi}$.                                      

   Another important scaling property is the $\beta $ function, defined as
\begin{eqnarray}
\beta (g)={\partial \log g \over{\partial \log L}}.
\end{eqnarray}
Fig.3(b) shows $\beta (g)$. 
$\beta$ is negative for small $g$ and approaches a positive
value for large $g$.
In the inset, $\beta$ is plotted as
a function of $\log g$ and we find that
$\beta \propto \log (g/g_{c})$ around
$\beta \simeq 0$, in common with other kinds of MIT\cite{Lee}.

 
\begin{figure}
\vspace{2mm}
\centering
\epsfig{file=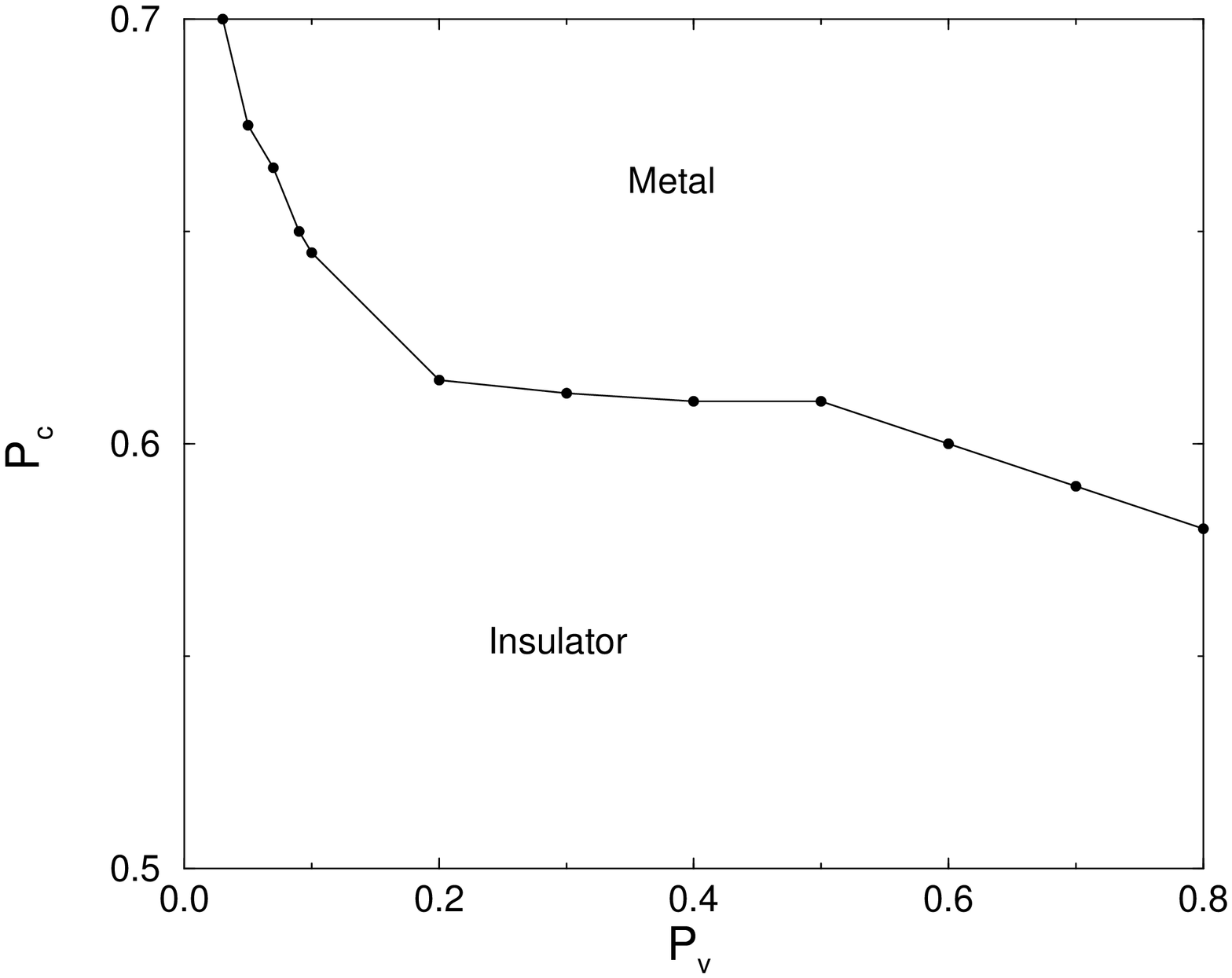, width=0.35\textwidth, angle=270}
\caption{
$P_{v}$ versus $P_{c}$.
\label{fig4}}
\end{figure}                                      

In our model, the phase coherence length $L_{\phi}$
is controlled by the probability
of voltage leads $P_{v}$ and the hopping between the a voltage lead and lattice
system $t_{v}$. Although it is difficult to determine the explicit
relation between $L_{\phi}$ and $P_{v}$ and $t_{v}$, it is
evident that for a fixed $t_{v}$, $L_{\phi}$ decreases with increasing
$P_{v}$ and approaches infinity (pure quantum system)
at $P_{v}=0$. We have carried out calculations
for several values of $P_{v}$ for a fixed $t_{v}=0.1$ to determine
the metal-insulator transition points.
Fig.4 presents the resulting MIT points $P_{c}$ versus $P_{v}$.


In the following, we make some remarks concerning our
results.

\noindent (i) We find that at the transition point $P_c$
the critical conductance $g_{c}$ is of the order
$e^{2}/h$. In our model $g_{c}$ depends on the on-site disorder $W$.
At $W=0$, $g_{c} \simeq 2e^{2}/h$, the conductance
for {\it one conducting channel} through the system with two
spin degeneracy\cite{Datta}. Thus,
we associate the name {\it meta-percolation} to the conduction process
at the transition point.

\noindent (ii) The meta-percolation proposed here is different
from the conventional classical percolation transition. Close to the
classical percolation point, the finite-size conductance scales
like $g(L)=L^{-x}f(L/\xi)$ with non-zero exponent $x$\cite{Mitescu}.
Thus, $g(L)$ for different $L$ will never cross at a single
point as for the case of meta-percolation shown in Fig.2(b).

\noindent (iii) The meta-percolation transition proposed in this work happens
at zero temperature as long as the phase coherence length
$L_{\phi}$ is finite. However, for a realistic electron system
where dephasing is caused by electron-electron or electron-phonon
interaction, $L_{\phi} \sim T^{-p}$ which diverges at $T=0$, and
in that case, the
meta-percolation MIT can only occur at finite temperatures.

Before ending, we comment on the possible connection between the meta-percolation
and the newly discovered MIT in 2D electron systems
\cite{Krav,DP_MIT,AlAs_elect,SiGe_holes,GaAs_holes,Simmons,Pudalov,Ribeiro,Dob,He,theories4,theories5,Belitz,theories6,theories11,theories7,theories9}.
It is well known that the dielectric constant of a 2D electron system
becomes negative at low density where
electron-electron interaction
is important. A consequence of the negative dielectric constant, 
we suggest, could be the formation of a droplet state.
The droplet state is a two-phase coexistence
region of high density electron liquid and low density electron ``gas". Here
we call it ``gas" purely for the reason that its density is low.
In fact, in the presence of impurities, the ``gas" phase is a disordered
Wigner crystal or a Wigner glass. 
The liquid phase has a better local conductivity compared to the ``gas" phase
because of its higher density. Thus, in the lattice model we have
studied, the liquid
phase corresponds to the region with nearest neighbor hopping
$t_{ij}=1$ while the "gas" phase corresponds to $t_{ij}=0$.
At finite temperatures, electron phase coherence length is finite, thus
the system is semi-quantum. In such a system, as we have shown, the
meta-percolation can induce MIT and recently discovered 2D MIT
may belong to this class. Unlike the true quantum phase transition,
this kind of MIT disappears at $T=0$ since electron phase coherence length
is infinity there. However, if dephasing mechanism can be maintained
at $T=0$, for instance, by adding magnetic impurities in the system,
then the meta-percolation MIT can survive even at zero temperature.

In summary, we have demonstrated that there is a novel type
percolation ({\it meta-percolation}) in two dimensional systems.
The meta-percolation can induce a metal-insulator transition
and the transition is semi-quantum in nature.

The work is supported by DOE 
under the contract number DE-FG03-98ER45687.



\begin{references}

\bibitem{Abrahams} E. Abrahams, P.W. Anderson, D.C. Licciardello,
and T.V. Ramakrishnan,
Phys. Rev. Lett. {\bf 42}, 673 (1979).

\bibitem{Lee} For a review of MIT, see for example,
P.A. Lee and T.V. Ramakrishnan, Rev. Mod. Phys. {\bf 57},
287 (1985).

\bibitem{Shapir} Y. Shapir, A. Aharony, and A.B. Harris,
Phys. Rev. Lett. {\bf 49}, 486 (1982); Y. Meir, A. Aharony, and
A.B. Harris, Europhys. Lett. {\bf 10}, 275 (1989);
Iksoo {\it et. al.}, Phys. Rev. Lett. {\bf 74}, 2094 (1995).

\bibitem{Buttiker} M. B\"{u}ttiker, Phys. Rev. B{\bf 33},
3020 (1986) and references therein.

\bibitem{Datta} S. Datta, {\it Electronic Transport in Mesoscopic
Systems} (Cambridge University Press, 1995).                     

\bibitem{Haug} H. Haug and A.P. Jauho, {\it Quantum Kinetics in
Transport and Optics of Semiconductors} (Springer 1996).

\bibitem{Shi} J.R. Shi and X.C. Xie, unpublished.

\bibitem{Barber} M.N. Barber, ``Phase Transitions and Critical
Phenomena'', Vol.8 (C. Domb and M.S. Green, eds),
Academic Press, London, p.145, 1983.

\bibitem{Kramer} A. Mackinnon and B. Kramer, Z. Physik, B {\bf 53}, 1 (1983).

\bibitem{Mitescu} For example, see C.D. Mitescu, {\em et al.}, J. Phys. A{\bf 15},
2523 (1982).

\bibitem{Krav} S. V. Kravchenko, {\em et al.},
Phys. Rev. B~{\bf 50}, 8039 (1994); S. V. Kravchenko, {\em et al.},
Phys. Rev. B~{\bf 51}, 7038 (1995); S. V. Kravchenko, {\em et al.},
Phys. Rev. Lett.~{\bf 77}, 4938 (1996); D. Simonian, {\em et al.},
Phys. Rev. Lett.~{\bf 79}, 2304 (1997).
 
\bibitem{DP_MIT} D. Popovi\'{c}, A. B. Fowler, and S. Washburn, Phys. Rev.
Lett.~{\bf 79}, 1543 (1997).
 
\bibitem{AlAs_elect} S. J. Papadakis and M. Shayegan, Phys. Rev. B~{\bf 57},
R15068 (1998).                                           

\bibitem{SiGe_holes} J. Lam, {\em et al.},
Phys. Rev. B~{\bf 56}, R12741 (1997); P. T. Coleridge, {\em et al.},
Phys. Rev. B~{\bf 56}, R12764 (1997).
 
\bibitem{GaAs_holes} Y. Hanein, {\em et al.},
Phys. Rev. Lett.~{\bf 80}, 1288 (1998); Y. Hanein, {\em et al.},
Phys. Rev. B{\bf 58}, R13338 (1998).

\bibitem{Simmons} M. Y. Simmons, {\em et al.},
Phys. Rev. Lett.~{\bf 80}, 1292 (1998); A.R. Hamilton, {\em et al.},
Phys. Rev. Lett.~{\bf 82}, 1542 (1999).                                 

\bibitem{Pudalov} V.M. Pudalov, {\em et al.}, JETP lett.~{\bf 66},
175 (1997).

\bibitem{Ribeiro} E. Riberio, {\em et al.}, Phys. Rev. Lett.~{\bf 82},
996 (1999).

\bibitem{Dob} V. Dobrosavljevi\'c, {\em et al.},
Phys. Rev. Lett.~{\bf 79}, 455(1997).
                                                                     
\bibitem{He} S. He and X. C. Xie, Phys. Rev. Lett.~{\bf 80}, 3324
(1998).
 
\bibitem{theories4} C. Castellani, {\em et al.},
Phys. Rev. B~{\bf 57}, R9381 (1998).
 
\bibitem{theories5} 
P. Phillips, {\em et al.},
Nature~{\bf 395}, 253 (1998). 

\bibitem{Belitz} D. Belitz and T. R. Kirkpatrick, Phys. Rev.
B~{\bf 58}, 8214 (1998).
 
\bibitem{theories6} Q. Si and C. M. Varma, Phys. Rev. Lett.~{\bf 81}, 4951
(1998).                                                                       

\bibitem{theories11} S. Chakravarty, {\em et al.},
preprint cond-mat/9805383 (1998).
                                                
\bibitem{theories7} B. L. Altshuler and D. L. Maslov, Phys. Rev. Lett.~{\bf
82}, 145 (1999).
 
\bibitem{theories9} T. M. Klapwijk and S. Das Sarma, Sol. St. Comm. (in press);
S. Das Sarma and E. H. Hwang, preprint cond-mat/9812216
(1998).
                                                                            
\end{references}
\end{document}